\documentclass[11pt]{article}
\usepackage{amssymb,latexsym,amsmath,amsbsy}
\usepackage[dvips]{graphicx}
\usepackage{cite}
\usepackage{hyperref}
\newtheorem{theorem}{Theorem}
\newtheorem{definition}[theorem]{Definition}

\newtheorem{remark}[theorem]{Remark}
\newtheorem{proposition}[theorem]{Proposition}
\DeclareMathOperator{\tr}{tr}
\newcommand{\myatop}[2]{\genfrac{}{}{0pt}{}{#1}{#2}}

\begin{document}
\begin{center}
{\Large \bf
A finite oscillator model with equidistant position spectrum\\[2mm] 
based on an extension of $\mathfrak{su}(2)$}\\[5mm]
{\bf Roy Oste, Joris Van der Jeugt} \\[1mm]
Department of Applied Mathematics, Computer Science and Statistics,\\
Ghent University, Krijgslaan 281-S9, B-9000 Gent, Belgium\\[1mm]
E-mail: Roy.Oste@UGent.be; Joris.VanderJeugt@UGent.be
\end{center}

\vskip 10mm

\noindent
Short title: finite oscillator for $\mathfrak{su}(2)_P$

\vskip 10mm

\begin{abstract}
We consider an extension of the real Lie algebra $\mathfrak{su}(2)$ by introducing a parity operator $P$ and a parameter $c$. 
This extended algebra is isomorphic to the Bannai-Ito algebra with two parameters equal to zero. 
For this algebra we classify all unitary finite-dimensional representations and show their relation with known representations of $\mathfrak{su}(2)$. 
Moreover, we present a model for a one-dimensional finite oscillator based on the odd-dimensional representations of this algebra. 
For this model, the spectrum of the position operator is equidistant and coincides with the spectrum of the known $\mathfrak{su}(2)$ oscillator. 
In particular the spectrum is independent of the parameter $c$ while the discrete position wavefunctions, which are given in terms of certain dual Hahn polynomials, do depend on this parameter.
\end{abstract}

\section{Introduction}
\label{sec1}

Finite oscillator models were introduced and investigated in a number of papers, see e.g.~\cite{Atak-Suslov, Atak2001, Atak2001b, Atak2005, JSV2011, JSV2011b}.
The standard and well-recognized example is the $\mathfrak{su}(2)$ oscillator model~\cite{Atak-Suslov,Atak2001}.
In brief, this model is based on the $\mathfrak{su}(2)$ algebra with basis elements $J_0=J_z$, $J_\pm=J_x\pm J_y$ satisfying
\begin{equation}
[J_0,J_\pm]=\pm J_\pm, \qquad [J_+,J_-]=2J_0,
\end{equation}
with unitary representations of dimension $2j+1$ (where $j$ is integer or half-integer). 
Recall that the oscillator Lie algebra can be considered as an associative algebra (with unit element $1$) with three generators $\hat H$, $\hat q$ and $\hat p$ (the Hamiltonian, the position and the momentum operator) subject to
\begin{equation}
[\hat H, \hat q] = -i \hat p, \qquad [\hat H,\hat p] = i \hat q, \qquad [\hat q, \hat p]=i,
\label{Hqp}
\end{equation}
in units with mass and frequency both equal to~1, and $\hbar=1$.
The first two are the Hamilton-Lie equations; the third the canonical commutation relation. 
The canonical commutation relation is not compatible with a finite-dimensional Hilbert space.
Following this, one speaks of a finite oscillator model if $\hat H$, $\hat q$ and $\hat p$ belong to some algebra such that the Hamilton-Lie equations are satisfied and such that the spectrum of $\hat H$ in representations of that algebra is equidistant~\cite{Atak2001,JSV2011}. 

In the $\mathfrak{su}(2)$ model, one chooses
\begin{equation}
\hat H=J_0+j+\frac12,\quad \hat q=\frac12(J_++J_-),\quad \hat p=\frac{i}{2}(J_+-J_-).
\label{su2-Hpq}
\end{equation}
These indeed satisfy $[\hat H, \hat q] = -i \hat p$, $[\hat H,\hat p] = i \hat q$, and in the representation $(j)$ labeled by $j$ the spectrum of $\hat H$ is equidistant (and given by $n+\frac12$; $n=0,1,\ldots,2j$).
Clearly, for this model the position operator $\hat q=\frac12(J_++J_-)$ also has a finite spectrum in the representation $(j)$ given by 
$q\in \{-j,-j+1,\ldots,+j\}$. 
In terms of the standard $J_0$-eigenvectors $|j,m\rangle$, the eigenvectors of $\hat q$ can be written as
\begin{equation}
|j,q) = \sum_{m=-j}^j \Phi_{j+m}(q) |j,m\rangle.
\label{4}
\end{equation}
The coefficients $\Phi_{n}(q)$ are the position wavefunctions, and in this model~\cite{Atak-Suslov,Atak2001} they turn out to be (normalized) symmetric Krawtchouk polynomials, $\Phi_n(q) \sim  K_ n(j+q;\frac12,2j)$.
The shape of the these wavefunctions is reminiscent of those of the canonical oscillator:
under the limit $j\rightarrow \infty$ they coincide with the 
canonical wavefunctions in terms of Hermite polynomials.

Following the ideas of the seminal papers on the $\mathfrak{su}(2)$ oscillator model, some alternative finite oscillator models were introduced~\cite{JSV2011, JSV2011b, JV2012}. 
The interest in these different models stems from several facts: in these new models additional parameters could be introduced, leading to wavefunctions with potentially more applications; the underlying algebras have a richer structure than $\mathfrak{su}(2)$; the wavefunctions are related to other classes of discrete orthogonal polynomials, and to new properties of these polynomials. 
In particular, we observed that in our models the wavefunctions were related to some ``doubling process'' of known orthogonal polynomials.
A peculiar property of the wavefunctions in the new models of Refs.~\cite{JSV2011, JSV2011b, JV2012}, which could be considered as a disadvantage, is that the support of the discrete position wavefunctions (which is the spectrum of the position operator) is no longer equidistant. 

So far, the introduction of new finite oscillator models looked rather arbitrarily. 
The mentioned relation to a ``doubling process'' for orthogonal polynomials, however, raised the question in how many ways the classical discrete orthogonal polynomials can be doubled, and whether these give rise to interesting models.
In a recent paper~\cite{DoubleHahn}, we investigated and classified all doubles for Hahn, dual Hahn and Racah polynomials, which are the standard discrete orthogonal polynomials one level up from the Krawtchouk polynomials in the Askey scheme~\cite{Koekoek}.
We not only classified all possible doubles; additionally we showed that each double is essentially a Christoffel-Geronimus pair~\cite{DoubleHahn}.

Following the classification of~\cite{DoubleHahn}, it is worthwhile to investigate the oscillator models corresponding to Hahn or dual Hahn doubles that have not yet been studied before. 
We are in particular interested in models in which also the position operator spectrum is equidistant.
This is how the present paper originated: from our classification~\cite{DoubleHahn} it is clear that there is one case (referred to as ``Dual Hahn~I'' in~\cite{DoubleHahn}) giving rise to a natural equidistant position spectrum.
The model related to this case is the subject here.

Rather than introducing this new model via the dual Hahn double, it is -- in the finite oscillator context -- more natural to start from the underlying algebra. 
This is the line followed here: in section~\ref{sec2} we introduce the algebra $\mathfrak{su}(2)_P$, an extension of $\mathfrak{su}(2)$ by a parity operator $P$. 
The extension is close to a ``central extension'', with parameter $c$, but $P$ is not a central element (it commutes with $J_0$ but anticommutes with $J_+$ and $J_-$). 
This algebra is interesting on its own, and we also classify all irreducible unitary finite-dimensional representations of $\mathfrak{su}(2)_P$.
These representations can be understood as deformations of the common $\mathfrak{su}(2)$ representations of dimension $2j+1$, except that not all of these can be deformed (which representations appear depends on the value of $c$, the parameter of the extension).
In section~\ref{sec3} we discuss the finite oscillator model related to $\mathfrak{su}(2)_P$. 
In particular, we show that (for odd dimensions) the spectral problem for the position operator is of type dual Hahn~I (according to the classification~\cite{DoubleHahn}), and we construct the orthonormal eigenvectors of the position and momentum operator.
The following section deals with some properties of the corresponding position wavefunctions.
The expressions of the wavefunctions are quite simple dual Hahn polynomials. 
We also discuss some plots of the wavefunctions, and state some natural limits (in particular to the canonical quantum oscillator).
The paper ends with some concluding remarks:
in particular, we clarify the connection/difference between the algebra $\mathfrak{su}(2)_P$ and previously used extended algebras $\mathfrak{u}(2)_\alpha$~\cite{JSV2011} and $\mathfrak{su}(2)_\alpha$~\cite{JSV2011b} in the context of ``Hahn oscillators'', and we discuss a reflection differential operator realization of $\mathfrak{su}(2)_P$.

\section{An extension of su(2) and its representations}
\label{sec2}

The real Lie algebra $\mathfrak{su}(2)$~\cite{Wybourne,Humphreys} can be defined by three basis elements 
$J_0$, $J_+$, $J_-$ with commutators $[J_0,J_\pm]=\pm J_\pm$ and $[J_+,J_-]=2J_0$.
The non-trivial unitary representations of $\mathfrak{su}(2)$, corresponding to the star relations $J_0^\dagger=J_0$, 
$J_\pm^\dagger=J_\mp$, are labelled~\cite{Wybourne,Humphreys} by a positive integer or half-integer $j$. 
These representations have dimension $2j+1$,
and the action on a set of basis vectors $|j,m\rangle$ (with $m=-j,-j+1,\ldots,+j$) is given by
\[
J_0 |j,m\rangle = m\,|j,m\rangle,\qquad 
J_\pm |j,m\rangle = \sqrt{(j\mp m)(j\pm m +1)}\;|j,m\pm 1\rangle.
\]

The Lie algebra $\mathfrak{su}(2)$ can be extended by a parity operator or involution $P$, whose action in these representations is given by
$P |j,m\rangle = (-1)^{j+m}\;|j,m\rangle$. 
On the algebraic level, this means that we extend the universal enveloping algebra of $\mathfrak{su}(2)$ by an operator $P$ that commutes with $J_0$, that {\em anticommutes} with $J_+$ and $J_-$, and for which $P^2=1$. 
Moreover, by means of this operator $P$ the standard $\mathfrak{su}(2)$ relations can be deformed introducing a real parameter $c$. 
This gives rise to an extension of the Lie algebra of $\mathfrak{su}(2)$ which itself is not a Lie algebra (nor a Lie superalgebra). 
This extension will be denoted by  $\mathfrak{su}(2)_P$ and is defined as follows.  
\begin{definition}
Let $c$ be a parameter. The algebra $\mathfrak{su}(2)_P$ is a unital algebra with basis elements $J_0$, $J_+$, $J_-$ and $P$ subject to 
the following relations:
	\begin{equation}
		\label{P}
		P^2=1, \qquad \lbrack P,J_0 \rbrack =PJ_0-J_0P= 0, \qquad \{P,J_{\pm} \} = PJ_{\pm} + J_{\pm}P  = 0 ,
	\end{equation}
and the $\mathfrak{su}(2)$ relations which are deformed as follows:
		\begin{align}
			\label{J0Jpm}
			\lbrack J_0,J_{\pm}\rbrack & = \pm J_{\pm} 
			\\
			\label{JpJm}
			\lbrack J_+,J_-\rbrack & = 2J_0 + cP\rlap{\,.}
		\end{align}
The star relation for this algebra is determined by:
\begin{equation}
P^\dagger=P, \qquad J_0^\dagger=J_0, \qquad J_\pm^\dagger=J_\mp.
\label{dagger}
\end{equation}
\end{definition}
For $c=0$ the deformed relation~\eqref{JpJm} reduces to the regular $\mathfrak{su}(2)$ relation. 
Note that this extension is very similar to a central extension; the only relation that violates this is the anticommutator in~\eqref{P}. 

The appearance of both a commutator and an anticommutator in~\eqref{P} also implies that one is not dealing with a Lie algebra nor with a Lie superalgebra.
The algebraic structure defined here is not new, however. 
The algebra $\mathfrak{su}(2)_P$ is in fact isomorphic to a special case of the Bannai-Ito algebra where two parameters are equal to zero\cite{DiracDunkl,BI}. 
Indeed, putting
\begin{equation}
K_1 = \frac{1}{2}(J_++J_-), \quad K_2 = -\frac{1}{2}(J_+-J_-)P, \quad K_3 = J_0P
\label{K1K2K3}
\end{equation}
we have
\[
\{K_1,K_2\} = K_3 + \frac{c}{2}, \quad  \{K_2,K_3\} = K_1 \quad  \{K_3,K_1\} = K_2\rlap{\,.}
\]
The star relations~\eqref{dagger} correspond to $K_i^\dagger=K_i$ for $i=1,2,3$. 
Moreover, this algebra can also be seen as a special case of the so-called algebra ${\cal H}$ of the dual $-1$ Hahn polynomials, see~\cite{Genest2013,Tsujimoto}, where one of the parameters equals zero. 
 
Using \eqref{J0Jpm} and \eqref{JpJm}, one easily shows that the Casimir element of $\mathfrak{su}(2)$, given by 
$\Omega = 2J_0^2 + J_+J_- + J_-J_+$, 
remains central for the universal enveloping algebra of $\mathfrak{su}(2)_P$. By means of \eqref{JpJm}, the Casimir element can also be written as
\begin{equation}
\label{Cas}
\Omega  =2J_+J_- + 2J_0^2  - 2J_0 - cP =2J_-J_+ + 2J_0^2  + 2J_0 + cP \rlap{\,.}
\end{equation}

Our purpose is now to determine all finite-dimensional unitary representations of $\mathfrak{su}(2)_P$, corresponding to the star conditions~\eqref{dagger}.

Let $(W,\rho_W)$ be a representation of $\mathfrak{su}(2)_P$. We consider $W$ as an $\mathfrak{su}(2)_P$ module by setting $G\cdot v = \rho_W(G)v$ for $G\in \mathfrak{su}(2)_P$ and $v\in W$. 
Take $v_0\in W$ to be an eigenvector of $J_0$ with eigenvalue $\lambda$. 
We will construct the $\mathfrak{su}(2)_P$ invariant subspace containing $v_0$. 
If $W$ is irreducible this space must be either $W$ or trivial. 
The trivial case results from $v_0$ being the zero vector, so from now on we assume that $v_0$ is not the zero vector.   

From \eqref{P} follows that $J_0Pv_0 =PJ_0v_0 = \lambda Pv_0 $, hence $Pv_0$ is also an eigenvector of $J_0$ with eigenvalue $\lambda$. 
We distinguish between two cases, $Pv_0$ is either a multiple of $v_0$ or not. 
If $Pv_0$ is linearly independent of $v_0$ and also has $\lambda$ as eigenvalue for $J_0$, the vectors $v^+_0 = v_0 + Pv_0$ and $v^-_0 = v_0 - Pv_0$ are also eigenvectors of $J_0$ and we have $Pv_0^+ = v_0^+$ and $Pv_0^- = -v_0^-$. 
The vectors $v_0^+$ and $v_0^-$ will then generate two different invariant subspaces, hence the representation $W$ is not irreducible. 
We may thus assume that $v_0$ is also an eigenvector of $P$.

If $J_0 v_0 =\lambda v_0$, then for a positive integer $k$, the vector $(J_{\pm})^k v_0$ is also an eigenvector of $J_0$. 
Indeed, using $\lbrack J_0,(J_{\pm})^k\rbrack = \pm k(J_{\pm})^k $, which follows from \eqref{J0Jpm}, we have 
\begin{equation}
\label{actJ0}
J_0 (J_{\pm})^k v_0 = \bigl( (J_{\pm})^k J_0 + \lbrack J_0,(J_{\pm})^k\rbrack \bigr) v_0  = J_{\pm} J_0 v_0  \pm k(J_{\pm})^k   J_{\pm} v_0 = (\lambda \pm k)  (J_{\pm})^k v_0 \rlap{\,.}
\end{equation}
Moreover, the vectors  $\bigl\{ (J_+)^k v_0 \,\big\vert\, k \in \mathbb{N} \bigr\}$ must be linearly independent because they have distinct eigenvalues as eigenvectors of $J_0$. 
If we impose $W$ to be finite-dimensional, then $(J_+)^k v_0=0$ for some $k\in \mathbb{N}$. 
Without loss of generality we may assume that $J_+v_0 = 0$, making $v_0$ the highest weight vector, i.e.~the eigenvector of $J_0$ with the highest eigenvalue, with corresponding highest weight $\lambda$. 

Following the same reasoning, the sequence $\bigl\{ (J_-)^k v_0 \,\big\vert\, k \in \mathbb{N} \bigr\}$ is also linearly independent and must terminate. 
We thus have $J_-(J_-^nv_0) = 0$ for some $n \in \mathbb{N}$ and we may assume without loss of generality that $n$ is minimal in this aspect, 
i.e.~$J_-^{n}v_0 \neq 0$. 
We will now show that the set
\begin{equation}
\label{basis}
\bigl\{ v_k =(J_-)^k v_0 \,\big\vert\, k = 0,\dots,n \bigr\}
\end{equation}
forms a basis for the $\mathfrak{su}(2)_P$ invariant subspace containing $v_0$. 
If $W$ is irreducible this space must be all of $W$. 
So far, we have \eqref{basis} being invariant under the action of $J_0$ and $J_-$, with $J_0 v_k=(\lambda-k) v_k$.
We now look at the action of $P$ and $J_+$ on \eqref{basis}. 
 
As $v_0$ is an eigenvector of $P$ and $P^2=1$, we necessarily have $
Pv_0 = \epsilon \, v_0
$ with $\epsilon = \pm 1$.  
Moreover, as $P$ anti-commutes with $J_-$ \eqref{P}, we find the action of $P$ on \eqref{basis} to be
\begin{equation}
\label{actP}
Pv_k = \epsilon (-1)^k v_k \rlap{\,.}
\end{equation}

For the action of $J_+$, we have $J_+v_0 = 0$, while for $k\geq 0$ we can write 
\[
J_+ v_{k+1} = J_+J_- v_{k} =\left( \frac12\Omega - J_0^2  + J_0 + \frac{c}{2}P \right)v_{k} 
\]
where $\Omega$ is the Casimir element \eqref{Cas} whose action is constant on $W$. 
Using $J_+v_0 = 0$ the action of $\Omega$ on $v_0$ is given by
\[
\Omega v_0 =  (2J_-J_+ + 2J_0^2  + 2J_0 + cP)v_0 =  (2\lambda^2 + 2\lambda + c\epsilon)v_0
\]
We thus find 
\begin{equation}
\label{Ak}
J_+ v_{k+1} = \left(   (k+1)(2\lambda-k) + c\, \epsilon\, \frac{1+(-1)^k}{2}
\right) v_{k} \equiv A(k) v_k \rlap{\,,}
\end{equation}
so \eqref{basis} forms the basis for a $\mathfrak{su}(2)_P$ invariant subspace.

Now, the value of the highest weight $\lambda$ follows from the action of $J_0$ and $P$ on the basis \eqref{basis}. 
Indeed, taking the trace of both sides of \eqref{JpJm} acting on $W$, we get
\begin{align}
0 = \tr\bigl(\lbrack J_+,J_-\rbrack\bigr)
=  2\tr(J_0) + \tr(cP) 
=  2(n+1)\lambda - n(n+1) + c\, \epsilon\, \frac{1+(-1)^n}{2} \rlap{\,.}
\label{lambda}
\end{align}
From which we find 
\begin{equation}
\label{hew}
\lambda =  \frac{n}{2} -  \frac{ c\, \epsilon}{2(n+1)}\frac{1+(-1)^n}{2}  \rlap{\,.}
\end{equation}
Note that the deformation parameter $c$ appears in the value of the highest weight only when $n$ is even, that is for odd-dimensional representations.

Substituting \eqref{hew} for $\lambda$ in \eqref{Ak} we arrive at
\begin{equation}
A(k)= \   (k+1) (n-k) + c\, \epsilon\, \left( \frac{1+(-1)^k}{2} - \frac{1+(-1)^n}{2} \frac{(k+1)}{(n+1)} \right)\rlap{\,.}
\end{equation}
For $n$ even, this reduces to 
\begin{equation}
\label{Akeven}
A(k)= \begin{cases}
(k+1)  \left(n-k  -  \frac{c\, \epsilon}{n+1} \right), & \text{if $k$ is odd;}\\
\left( k+1 +  \frac{c\, \epsilon}{n+1} \right)(n-k), & \text{if $k$ is even,}
\end{cases}
\end{equation}
while for $n$ odd
\begin{equation}
\label{Akodd}
A(k)= \begin{cases}
(k+1) (n-k), & \text{if $k$ is odd;}\\
(k+1) (n-k) +c\, \epsilon , & \text{if $k$ is even.}
\end{cases}
\end{equation}

Next, we require that the representation $W$ is unitary under the star conditions~\eqref{dagger}. 
Hereto, we introduce a sesquilinear form $\langle \cdot,\cdot \rangle \colon W \times W \to \mathbb{C}$ such that
\[
\langle v_k, v_{\ell} \rangle = h_k\,  \delta_{k,\ell}  \rlap{\,,}
\]
where we can put $h_0=1$ or $\langle v_0, v_0 \rangle = 1 $.
In order to be an inner product we need $h_k>0$ for $k\geq 0$. 
For $k\geq 1$ we have, imposing the star condition $J_-^{\dagger}= J_+$, 
\begin{equation}
\label{hk}
	h_{k} = \langle v_{k}, v_{k} \rangle = \langle J_- v_{k-1},  v_{k} \rangle  = \langle  v_{k-1}, J_+ v_{k} \rangle = A(k-1)  \langle  v_{k-1}, v_{k-1} \rangle = A(k-1) h_{k-1}  \rlap{\,.}
\end{equation}
This is strictly positive if $A(k)>0$ for $0 \leq k\leq n-1$. 
Distinguishing between $n$ even and odd, we find that 
\eqref{Akeven} is strictly positive for $-(n+1) < c\, \epsilon  < n+1$, while 
\eqref{Akodd} is strictly positive if $ c\, \epsilon >-n  $. 

The star conditions $P^\dagger=P$, $J_0^\dagger=J_0$ are satisfied as $P$ and $J_0$ have real eigenvalues on $v_k$. 
Putting $j=n/2$ and introducing the orthonormal basis
\[
  |j,m\rangle = \frac{v_{j-m}}{\lVert v_{j-m} \rVert} \qquad (m = -j, -j+1, \dots , j-1, j)
\]
where $\lVert v_k \rVert = \sqrt{ \langle  v_{k}, v_{k} \rangle }  = \sqrt{ h_{k} } $, we find using \eqref{hk}
\[
J_- |j,m \rangle = J_- \frac{v_{j-m}}{ \lVert v_{j-m} \rVert} = \frac{v_{j-m+1}}{ \sqrt{ h_{j-m} }} = \sqrt{ A(j-m) }  |j,m-1 \rangle
\]
and
\[
J_+ | j,m \rangle = J_+ \frac{v_{j-m}}{\lVert v_{j-m} \rVert}  = A(j-m-1) \frac{v_{j-m-1}}{ \sqrt{ h_{j-m} }}  = \sqrt{ A(j-m-1) }  |j,m+1 \rangle \rlap{\,.}
\]
We summarize this in the following result: 

\begin{proposition}\label{prop1}
For a given real parameter $c$ and choice of $\epsilon = \pm 1$, we have the following irreducible unitary finite-dimensional representations of $\mathfrak{su}(2)_P$, corresponding to the star conditions~\eqref{dagger}: 
	
For every positive integer $j$ such that $2j+1 > \lvert c \rvert$, we have an odd-dimensional representation of dimension $2j+1$. 
The action of the $\mathfrak{su}(2)_P$ operators on a set of basis vectors 
	$|j,-j\rangle$, $|j,-j+1\rangle$, $\ldots$, $|j,j\rangle$ is given by:
	\begin{align}
	& P |j,m\rangle =  \epsilon (-1)^{j+m}\;|j,m\rangle, \label{act-Pe}\\
	& J_0 |j,m\rangle = (m -\tilde{c} /2) \;|j,m\rangle,\label{act-J0e}\\
	& J_+ |j,m\rangle = 
	\begin{cases}
	\sqrt{(j-m+  \tilde{c})(j+m +1 )}\;|j,m+1\rangle, & \text{if $j+m$ is odd;}\\
	\sqrt{(j-m)(j+m+1- \tilde{c})}\;|j,m+1\rangle, & \text{if $j+m$ is even.}
	\end{cases} \label{act-J+e}\\
	& J_- |j,m\rangle = 
	\begin{cases}
	\sqrt{(j+m - \tilde{c})(j-m+1)}\;|j,m-1\rangle, & \text{if $j+m$ is odd;}\\	
	\sqrt{(j+m)(j-m +1 +  \tilde{c})}\;|j,m-1\rangle, & \text{if $j+m$ is even.}
	\end{cases} \label{act-J-e}
	\end{align} 
	where $\tilde{c} = c\,\epsilon/(2j+1)$. Note that $2j+1 > \lvert c \rvert$ is equivalent to $|\tilde c|<1$.
	
For every positive half-integer $j$  such that 	$ 2j > - c\, \epsilon $, we have an even-dimensional representation of dimension $2j+1$. 
The action of the $\mathfrak{su}(2)_P$ operators on a set of basis vectors 
	$|j,-j\rangle$, $|j,-j+1\rangle$, $\ldots$, $|j,j\rangle$ is given by:
	\begin{align}
	& P \: |j,m\rangle = \epsilon (-1)^{j+m+1}\;|j,m\rangle, \label{act-P}\\
	& J_0 \, |j,m\rangle = m\;|j,m\rangle,\label{act-J0}\\
	& J_+  |j,m\rangle = 
	\begin{cases}
	\sqrt{(j-m)(j+m+1)}\;|j,m+1\rangle, & \text{if $j+m$ is odd;}\\
	\sqrt{(j-m)(j+m+1)+c\, \epsilon}\;|j,m+1\rangle, & \text{if $j+m$ is even.}
	\end{cases} \label{act-J+}\\
	& J_-  |j,m\rangle = 
	\begin{cases}
	\sqrt{(j+m)(j-m+1)+c\, \epsilon}\;|j,m-1\rangle, & \text{if $j+m$ is odd;}	\\
	\sqrt{(j+m)(j-m+1)}\;|j,m-1\rangle, & \text{if $j+m$ is even.}
	\end{cases} \label{act-J-}
	\end{align} 
\end{proposition}

We can write the actions of $J_+$ and $J_-$ in the above result more compactly. 
For $j$ an integer
\[
J_{\pm} |j,m\rangle = 
\begin{cases}
\sqrt{(j\mp m\pm  \tilde{c})(j\pm m +1 )}\;|j,m\pm 1\rangle, & \text{if $j+m$ is odd;}\\
\sqrt{(j\mp m)(j\pm m+1\mp \tilde{c})}\;|j,m\pm 1\rangle, & \text{if $j+m$ is even;}
\end{cases} 
\]
while for $j$ a half-integer
\[
J_{\pm}  |j,m\rangle = 
\begin{cases}
\sqrt{(j\mp m)(j\pm m+1)}\;|j,m\pm 1\rangle, & \text{if $j\pm m$ is odd;}\\
\sqrt{(j\mp m)(j\pm m+1)+c\, \epsilon}\;|j,m\pm 1\rangle, & \text{if $j\pm m$ is even.}
\end{cases}
\]
The action of the Casimir \eqref{Cas} is indeed scalar on these representations, and given by
\[
\bigl(2J_0^2 + J_+J_- + J_-J_+\bigr) |j,m\rangle  =
\begin{cases}
2j(j+1)+c\, \epsilon & \text{if $j$ is a half-integer;}\\
2j(j+1)+\frac{\tilde{c}^2}{2} & \text{if $j$ is an integer,}
\end{cases} 
\]
again with $\tilde{c} = c\,\epsilon/(2j+1)$.

\begin{remark}
For $j$ a half-integer, these representations correspond precisely to those of the unital algebra $\mathfrak{u}(2)_\alpha$ \cite{JSV2011}, which contains moreover an extra central operator $C$ with diagonal action $C |j,m\rangle=(2j+1)|j,m\rangle$.  
Indeed, substituting $c\, \epsilon = (2\alpha +1)^2 + (2\alpha +1)(2j+1)$ we find the same action as in \cite{JSV2011}:
\begin{align*}
&	(j-m)(j+m+1)+ (2\alpha +1)^2 + (2\alpha +1)(2j+1) = 	(j-m + 2\alpha +1 )(j+m+2\alpha +2), \\
&	(j+m)(j-m+1)+ (2\alpha +1)^2 + (2\alpha +1)(2j+1) = 	(j-m + 2\alpha +2 )(j+m+2\alpha +1).
\end{align*}
\end{remark}

For this reason, only the representations with $j$ integer are new in the context of finite oscillator models.
And therefore, only the odd-dimensional representations will play a role in the following sections.

\begin{remark} 
In \cite{DiracDunkl} the finite-dimensional unitary representations of the Bannai-Ito algebra~\cite{GVZ2014,GVZ2014b} corresponding to a realization in terms of Dirac-Dunkl operators were determined (see also~\cite{Huang} for a more general approach). 
Since the algebra $\mathfrak{su}(2)_P$ is isomorphic to a special case of the Bannai-Ito algebra, we should observe a correspondence between the representations in Proposition~\ref{prop1} and those of~\cite{DiracDunkl}.
One difference, however, is that in our case the parameter $c$ is the basic parameter, and its value determines the existence of representations of certain dimensions.  

Returning to~\cite{DiracDunkl}, the general Bannai-Ito algebra with three parameters $\omega_1,\omega_2,\omega_3$ is characterized by three real numbers $\mu_1,\mu_2,\mu_3$ appearing in the Dunkl operators, and a positive integer $N$ (with $N+1$ the dimension of the representation).
For even $N$, i.e.\ $N=2j$ with $j$ integer, the following choice for $\mu_i$:
\begin{equation}
\mu_1=\mu_2=-\frac{N+1+\tilde c}{4}, \qquad \mu_3=-\frac{N+1-\tilde c}{4}
\label{mui}
\end{equation}
in \cite[eq.~(48)]{DiracDunkl} leads to the same (matrix) representation for $K_1, K_2, K_3$ as our representation~\eqref{act-Pe}--\eqref{act-J-e} used in~\eqref{K1K2K3}.
Note that with $|\tilde c|<1$, the above $\mu_i$-values are negative. Strictly speaking, only nonnegative values for $\mu_i$ were considered in~\cite{DiracDunkl}. It is clear, however, that for the values~\eqref{mui} the matrix elements $U_k$ appearing in \cite[eq.~(48)]{DiracDunkl} are still positive and thus these values are also allowed.

For odd $N$, i.e.\ $N=2j$ with $j$ half-integer, the correspondence is not so simple.
For that case, the basis vectors $|N,k\rangle$ of \cite{DiracDunkl} are not the same as our basis vectors $|j,m\rangle$ for a particular choice of $\mu_1,\mu_2,\mu_3$. 
So the correspondence between the matrix representations becomes complicated and we do not include it here.
\end{remark}

\section{A one-dimensional oscillator model} 
\label{sec3}

We now consider a model for a one-dimensional finite oscillator based on the odd-dimensional representations of the algebra $\mathfrak{su}(2)_P$, that is for $j$ an integer. 
We will see that for this model the spectrum of the position operator is independent of the parameter $c$, equidistant and coincides with the spectrum of the $\mathfrak{su}(2)$ oscillator~\cite{Atak2001}. 
The eigenvectors (and thus also the position wavefunctions) do depend on the additional parameter $c$.

Following the notation and ideas of section~\ref{sec1}, we have to choose a position, 
momentum and Hamiltonian operator ($\hat q$, $\hat p$, $\hat H$) from the algebra $\mathfrak{su}(2)_P$ such that the Hamilton-Lie equations are satisfied, and such that the spectrum of $\hat H$ in a representation is equidistant.
Given $\mathfrak{su}(2)_P$ with parameter $c$, and the representation of dimension $2j+1$ ($j$ integer) determined in Proposition~\ref{prop1}, 
with $2j+1>|c|$ and $\epsilon=1$, the following choice is natural and follows~\cite{JSV2011,JSV2011b}:
\begin{equation}
\hat q = \frac12 (J_++J_-), \qquad
\hat p = \frac{i}{2}(J_+-J_-), \qquad
\hat H = J_0+j+\frac{\tilde{c}}{2} +\frac12,
\end{equation}
where $\tilde c = c/(2j+1)$ and thus the parameter $\tilde c$ satisfies $-1<\tilde c<+1$.

It is easy to verify that the first two equations of~\eqref{Hqp} are satisfied and moreover from \eqref{act-J0e} it follows that on $|j,m\rangle$ the spectrum of $\hat H$ is indeed linear and given by
\begin{equation}
n+\frac12 \qquad (n=0,1,\ldots,2j).
\end{equation}

From the actions~\eqref{act-J+e}--\eqref{act-J-e}, one finds for even $j+m$
\begin{equation*}
2\hat q |j,m\rangle = \sqrt{(j+m)(j-m +1 +  \tilde{c})}\;|j,m-1\rangle +
\sqrt{(j-m)(j+m+1- \tilde{c})}\;|j,m+1\rangle\rlap{\,,}
\end{equation*}
while for odd $j+m$ 
\begin{equation*}
2\hat q |j,m\rangle = 	\sqrt{(j+m - \tilde{c})(j-m+1)}\;|j,m-1\rangle +
\sqrt{(j-m+  \tilde{c})(j+m +1 )}\;|j,m+1\rangle\rlap{\,.}
\end{equation*}
The action of $2i\hat p$ is similar. 
For the representation space, denoted here by $W_j$, we choose the following (ordered) basis:
\begin{equation}
\{ |j,-j\rangle, |j,-j+1\rangle, \ldots, |j,j-1\rangle, |j,j\rangle \},
\label{B}
\end{equation}
and then the operators $2\hat q$, $2i\hat p$ take the matrix forms
\begin{equation}
2\hat q=\left(
\begin{array}{ccccc}
0 & M_0& 0 & \cdots & 0 \\
M_0 & 0 & M_1 & \cdots & 0\\
0 & M_1 & 0 & \ddots &  \\
\vdots & \vdots & \ddots & \ddots& M_{2j-1}\\
0 & 0 &  & M_{2j-1} & 0
\end{array}
\right) \equiv M^q,
\label{Mq}
\end{equation}
\begin{equation}
2i\hat p=\left(
\begin{array}{ccccc}
0 & M_0& 0 & \cdots & 0 \\
-M_0 & 0 & M_1 & \cdots & 0\\
0 &- M_1 & 0 & \ddots &  \\
\vdots & \vdots & \ddots & \ddots& M_{2j-1}\\
0 & 0 &  & -M_{2j-1} & 0
\end{array}
\right)\equiv M^p,
\label{Mp}
\end{equation}
with
\begin{equation}
M_k= \begin{cases}
\sqrt{(k+1- \tilde{c})(2j-k)},& \text{if $k$ is even;}\\
\sqrt{(k+1)(2j-k+\tilde{c})},  & \text{if $k$ is odd.}
\end{cases}
\label{Ma}
\end{equation}
For these matrices, the eigenvalues and eigenvectors are known explicitly: 
the system is of type ``dual Hahn~I''~\cite[Proposition~2]{DoubleHahn} (with $\gamma+\delta+1=0$ in the notation of~\cite{DoubleHahn}). 
The expressions of the eigenvectors involve dual Hahn polynomials, so let us first recall some notation.

For a positive integer $N$, the dual Hahn polynomial of degree $n$ ($n=0,1,\dots,N)$ in the variable $\lambda(x)=x(x+\gamma+\delta+1)$, with parameters $\gamma>-1$ and $\delta>-1$ (or $\gamma < -N$ and $\delta <-N$) is defined by~\cite{Koekoek,Suslov,Ismail}:
\begin{equation}\label{DHahn}
R_n(\lambda(x);\gamma,\delta,N) =  
{\;}_3F_2\left(\myatop{-x,x+\gamma+\delta+1,-n}{\gamma+1,-N};1 \right)
\end{equation}
in terms of the generalized hypergeometric series $_3F_2$ of unit argument~\cite{Bailey,Slater}. 
Dual Hahn polynomials satisfy a (discrete) orthogonality relation~\cite{Koekoek}:
\begin{equation}
\sum_{x=0}^N w(x;\gamma, \delta,N) R_n(\lambda(x);\gamma, \delta, N) R_{n'}(\lambda(x);\gamma, \delta, N) = h_n(\gamma,\delta,N)\, \delta_{n,n'},
\label{orth-R}
\end{equation} 
where
\begin{align}
& w(x;\gamma, \delta,N) = \frac{(2x+\gamma+\delta+1)(\gamma+1)_x(N-x+1)_x N!}{(x+\gamma+\delta+1)_{N+1}(\delta+1)_x x!} \qquad (x=0,1,\ldots,N),\nonumber\\
& h_n(\gamma, \delta,N)= \left[\binom{\gamma+n}{n} \binom{N+\delta-n}{N-n}\right]^{-1} . \label{wh}
\end{align}
We have used here the common notation for Pochhammer symbols~\cite{Bailey,Slater}
$(a)_k=a(a+1)\cdots(a+k-1)$ for $k=1,2,\ldots$ and $(a)_0=1$.
As $w$ is the weight function and $h_n(\gamma, \delta,N)$ the ``squared norm'', orthonormal dual Hahn functions $\tilde R$ are determined by:
\begin{equation}
\tilde R_n(\lambda(x);\gamma,\delta,N) \equiv \frac{\sqrt{w(x;\gamma,\delta,N)}\, R_n(\lambda(x);\gamma,\delta,N)}{\sqrt{h_n(\gamma,\delta,N)}}.
\label{R-tilde}
\end{equation}

From~\cite[Proposition~2]{DoubleHahn}, using the substitution $\gamma=(-\tilde c -1)/2$ and $\delta = (\tilde c-1)/2 = -\gamma-1$, we have:
\begin{proposition}
\label{propU}
The $2j+1$ eigenvalues of the position operator $\hat q$ in the representation $W_j$ are given by
		\begin{equation}
		\label{ev}
		-j, -j+1, \ldots,-1,0, 1, \ldots,j-1,j\rlap{\,.}
		\end{equation}
	The orthonormal eigenvector of the position operator $\hat q$ in $W_j$ for the eigenvalue $q$, denoted by $|j,q)$, is given 
	in terms of the basis~\eqref{B} by
	\begin{equation}
	|j,q) = \sum_{m=-j}^j U_{j+m,j+q} |j,m\rangle.
	\end{equation}
	Herein, $U=(U_{kl})_{0\leq k,l \leq 2j}$ is the $(2j+1)\times(2j+1)$ matrix with elements 
	\begin{align}
 &U_{2r,j} = (-1)^r \tilde R_r(\lambda(0);(-\tilde c -1)/2,(\tilde c -1)/2,j),\qquad U_{2r+1,j}=0, \nonumber\\
 &U_{2r,j-s} = U_{2r,j+s} = \frac{(-1)^r}{\sqrt{2}} \tilde R_r(\lambda(s);(-\tilde c -1)/2,(\tilde c -1)/2,j), \quad(r\in\{0,\ldots,j\};\; s\in\{1,\ldots,j\}); \label{Ueven} \\
 &U_{2r+1,j-s-1} = -U_{2r+1,j+s+1} = -\frac{(-1)^r}{\sqrt{2}} \tilde R_{r}(\lambda(s);(1-\tilde c)/2,(\tilde c +1)/2,j-1), 
\quad(r,s\in\{0,\ldots,j-1\})\label{Uodd} 
	\end{align}	
	where the functions $\tilde R$ are normalized dual Hahn polynomials~\eqref{R-tilde}.
	
The matrix $U$ is an orthogonal matrix, $UU^T=U^TU=I$, hence the $\hat q$ eigenvectors are orthonormal:
\[
(j, q | j, q') = \delta_{q,q'}.
\]Moreover, 
	\[M^q U = U D^q ,\]
	where $D^q$ is a diagonal matrix containing the eigenvalues \eqref{ev}.
\end{proposition}
This is the same spectrum as that of $\hat q$ in the $\mathfrak{su}(2)$ oscillator model~\cite{Atak2001}. 
For the latter model the eigenvectors could be expressed in terms of the Krawtchouk orthogonal polynomials. 
Now, the eigenvectors of the position operator have components proportional to dual Hahn polynomials with parameters $(-\tilde c -1)/2$ and $(\tilde c -1)/2$ when the component has even index, and with parameters $(1-\tilde c)/2$ and $(1+\tilde c)/2$ when the component has odd index. 
With the condition $|\tilde c|<1$ (see Proposition~\ref{prop1}), the weight functions of the dual Hahn polynomials are positive.

The matrix $M^p$ of the momentum operator $\hat{p}$ is up to signs the same as the matrix $M^q$. It has the same spectrum \eqref{ev} and for the eigenvectors we have the following result.
\begin{proposition}	
\label{propV}
	The orthonormal eigenvector of the momentum operator $\hat p$ in $W_j$ for the eigenvalue $p$, denoted by $|j,p)$, is given 
	in terms of the basis~\eqref{B} by
	\begin{equation}
	|j,p) = \sum_{m=-j}^j V_{j+m,j+p} |j,m\rangle.
	\end{equation}
Herein, $V=(V_{rs})_{0\leq r,s \leq 2j}$ is the unitary $(2j+1)\times(2j+1)$-matrix, $VV^{\dagger} =V^{\dagger}V =  I$,  defined by
	\[
	V = \mathcal{J} U,
	\]
	where $\mathcal{J}=-i\, \mathrm{diag}(i^0,i^1,i^2,\dots,i^{2j})$ and $U$ is the matrix determined in Proposition~\ref{propU}. 
\end{proposition}

\begin{remark}
The recurrence relation for the pair of polynomials appearing in \eqref{Ueven} and \eqref{Uodd} comes from $M^qU=U D^q$. 
By the form~\eqref{Mq}, this recurrence relation has zero diagonal term.
This is because the corresponding polynomials can be seen as an example of Chihara's construction~\cite[Section~8]{Chihara} of symmetric orthogonal polynomials, but applied to discrete orthogonal polynomials.
\end{remark}

\section{Oscillator wavefunctions and their properties}
\label{sec4}

The position (resp.~momentum) wavefunctions are the overlaps between the normalized eigenstates of the position operator $\hat q$ (resp.~the momentum operator $\hat p$) and the eigenstates of the Hamiltonian.
So the wavefunctions of the $\mathfrak{su}(2)_P$ finite oscillator are the overlaps between the $\hat q$-eigenvectors
and the $\hat H$-eigenvectors (or equivalently, the $J_0$-eigenvectors $|j,m\rangle$).
We will denote the position wavefunctions by $\phi^{(c)}_{j+m}(q)$ and the momentum wavefunctions by $\phi^{(c)}_{j+m}(q)$, where $m$, $q$ and $p$ assume one of the 
discrete values $-j,-j+1,\ldots,+j$. 
Concretely, following the notation of the previous section:
\begin{align}
\phi^{(c)}_{j+m}(q)  & =   \langle j,m | j,q ) = U_{j+m,j+q}, \\
\psi^{(c)}_{j+m}(p)  & =   \langle j,m | j,p ) = V_{j+m,j+p}.
\end{align}
Let us examine the explicit form of these functions in more detail, first for the position variable. 
The index $j+m$ ranges from $0$ to $2j$. For $j+m$ even, say $j+m=2n$, $\phi^{(c)}_{2n}(q)$ is by~\eqref{Ueven} an even function of the position variable $q$. For $ q=-j, -j+1, \ldots, j$ we have
\begin{equation}
\phi^{(c)}_{2n} (q) = \frac{(-1)^n}{\sqrt{2-\delta_{q,0}}} 
\sqrt{W(n,  q;\tilde c ,j)} \
{\;}_3F_2\left(\myatop{-q,q,-n}{(1-\tilde c)/2,-j};1 \right)
\label{Phi-even}
\end{equation}
where 
\[
W(n,  q;\tilde c ,j) = {\frac{w(|q|;(-\tilde c -1)/2,(\tilde c -1)/2,j)}{h_n((-\tilde c -1)/2,(\tilde c -1)/2,j)}} \rlap{\,,}
\]
with $w$ and $h_n$ as in \eqref{wh}. 
For $j+m$ odd, say $j+m=2n+1$, it is by~\eqref{Uodd} an odd function of the variable $q$. For $ q=-j, -j+1, \ldots, j$ we have
\begin{equation}
\phi^{(c)}_{2n+1} (q) = (-1)^n 
q \frac{\sqrt{(2n+1-\tilde{c})(j-n)}}{(1-\tilde{c})j}\sqrt{W(n,  q;\tilde c ,j)}\ 
{\;}_3F_2\left(\myatop{-q+1,q+1,-n}{(3-\tilde c)/2,-j+1};1 \right) \rlap{\,,}
\label{Phi-odd}
\end{equation}
where we used
\[
\frac{w(|q|;(1-\tilde c)/2,(\tilde c +1)/2,j-1)}{h_n((1-\tilde c)/2,(\tilde c +1)/2,j-1)} = q^2\frac{2(2n+1-\tilde{c})(j-n)}{(1-\tilde{c})^2j^2} W(n,  q;\tilde c ,j) \rlap{\,.}
\]
For the momentum wavefunctions we find in exactly the same manner
\begin{equation}
\psi^{(c)}_{2n} (p) = \frac{-i}{\sqrt{2-\delta_{p,0}}} 
\sqrt{W(n,  p;\tilde c ,j)} \
{\;}_3F_2\left(\myatop{-p,p,-n}{(1-\tilde c)/2,-j};1 \right)
\label{Psi-even}
\end{equation}
\begin{equation}
\psi^{(c)}_{2n+1} (p) =  
p \frac{\sqrt{(2n+1-\tilde{c})(j-n)}}{(1-\tilde{c})j}\sqrt{W(n,  p;\tilde c ,j)}\ 
{\;}_3F_2\left(\myatop{-p+1,p+1,-n}{(3-\tilde c)/2,-j+1};1 \right)\rlap{\,.}
\label{Psi-odd}
\end{equation}

\begin{remark}
Before we examine the behaviour of these discrete wavefunctions, let us comment on the distinction with the closely related dual $-1$ Hahn polynomials considered in~\cite{Tsujimoto}.
For this purpose, let us compare the polynomial expressions in~\eqref{Phi-even}--\eqref{Phi-odd}, i.e.\
\[
{\;}_3F_2\left(\myatop{-q,q,-n}{(1-\tilde c)/2,-j};1 \right), \qquad q\times{\;}_3F_2\left(\myatop{-q+1,q+1,-n}{(3-\tilde c)/2,-j+1};1 \right)
\]
with equations~(4.6) and (4.7) from~\cite{Tsujimoto}, in which one puts $N=2j$, i.e.\
\[
{\;}_3F_2\left(\myatop{-\frac{x}{4}+\eta,\frac{x}{4}+\eta ,-n}{1-\frac{\alpha}{2},-j};1 \right), \qquad 
(\frac{x}{4}-j-\eta){\;}_3F_2\left(\myatop{-\frac{x}{4}+\eta,\frac{x}{4}+\eta,-n}{1-\frac{\alpha}{2},-j+1};1 \right).
\]
For a particular choice of $\eta$ and $\alpha$, the even polynomials coincide, but the odd polynomials do not.
The reason is that the dual Hahn double of this paper corresponds to a Christoffel-Geronimus pair with parameter $\nu=0$ 
(see \cite[Section~5]{DoubleHahn}) and are of type ``dual Hahn~I'' in the terminology of~\cite{DoubleHahn}, 
whereas the dual $-1$ Hahn polynomials seem to correspond to a Christoffel transform for dual Hahn polynomials with a different parameter $\nu=j$, 
and are of type ``dual Hahn~II'' in the terminology of~\cite{DoubleHahn}.
\end{remark}

It is interesting to study these discrete wavefunctions for varying values of $\tilde c$, $-1<\tilde c<1$. 
For the special value $\tilde{c}=0$, the algebra $\mathfrak{su}(2)_P$ reduces to $\mathfrak{su}(2)$ and it is known that in this case, the wavefunctions $\phi^{(0)}_n(q)$ are in fact Krawtchouk functions. 
Indeed, when $\tilde{c}=0$ the dual Hahn polynomials, which are ${}_3F_2$ series
appearing in~\eqref{Phi-even}--\eqref{Phi-odd}, reduce to ${}_2F_1$ series
according to
\begin{align}
& {\;}_3F_2\left(\myatop{-q,q,-n}{1/2,-j};1 \right) = (-1)^n \frac{\binom{2j}{2n}}{\binom{j}{n}} 
{\;}_2F_1\left(\myatop{-2n,-j-q}{-2j};2 \right),\\
& {\;}_3F_2\left(\myatop{-q+1,q+1,-n}{3/2,-j+1};1 \right) = -\frac{(-1)^n}{2q} \frac{\binom{2j}{2n+1}}{\binom{j-1}{n}} 
{\;}_2F_1\left(\myatop{-2n-1,-j-q}{-2j};2 \right) \rlap{\,.}
\end{align}
These reductions have been given in~\cite{SV2011} and can be obtained, e.g., from~\cite[(48)]{Atak2005}. 
The ${}_2F_1$ series in the right hand side correspond to symmetric Krawtchouk polynomials
(i.e.\ Krawtchouk polynomials with $p=1/2$~\cite{Koekoek}). 
When $j$ tends to infinity, they yield the canonical oscillator wavefunctions~\cite{Atak2005} in terms of Hermite polynomials. 

\begin{figure}[htb]
\includegraphics{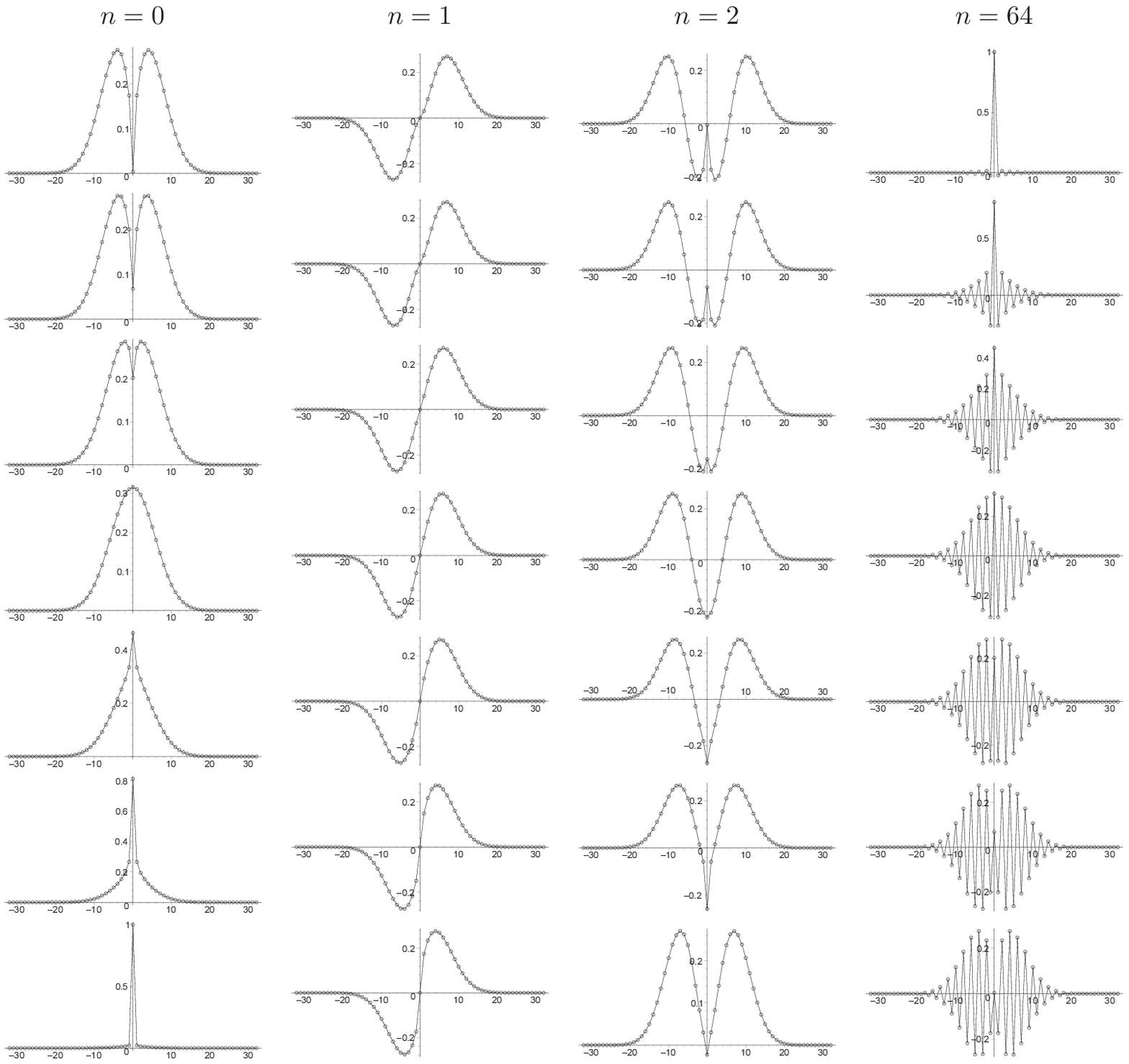}
\caption{Plots of the discrete wavefunctions $\phi^{(c)}_n(q)$ in the representation with $j=32$
	for the values $\tilde c=-0.999$, $\tilde c=-0.8$, $\tilde c=-0.3$, $\tilde c=0$, $\tilde c=0.3$, $\tilde c=0.8$, $\tilde c=0.999$ from top to bottom.
	The wavefunctions are plotted for $n=0,1,2$ and $n=64$.}
	\label{fig1}
\end{figure}

To investigate what happens for other values of $\tilde{c}$ we now choose a fixed value of $j$, namely $j=32$, and plot some of the 
wavefunctions $\phi^{({c})}_n(q)$ for various values of $\tilde{c}$. Recall (Proposition~\ref{prop1}) that $-1 < \tilde{c} < 1$ in order to have a unitary irreducible representation. 
In Figure~\ref{fig1} we take the following values for $\tilde c$, respectively,
\[
-0.999, \quad -0.8, \quad -0.3, \quad 0,\quad 0.3, \quad 0.8,\quad 0.999 \;.
\]
We also plot in each case the ground state $\phi^{(c)}_0(q)$ (left column), some low energy states $\phi^{(c)}_1(q)$ and $\phi^{(c)}_2(q)$ (2nd and 3rd column), and the highest energy state $\phi^{(c)}_{64}(q)$ (4th column).

Particularly interesting behaviour is observed when $\tilde{c}$ approaches the boundary values $-1$ or $1$. 
These bounds correspond to the disallowed value $-1$ for one of the parameters $\gamma$ or $\delta$ in the dual Hahn polynomial~\eqref{DHahn}. 
When $\tilde c$ tends to $-1$, the components of the highest energy state all tend to zero except for $q=0$ which tends to $1$. 
For all the other states, the value at $q=0$ tends to $0$.
When $\tilde c$ tends to $+1$, it is for the lowest energy state that all components tend to zero and the component at $q=0$ goes to $1$.
Similarly as for the other limit, for all the other states, the value at $q=0$ tends to $0$.
It can be verified that in these limits for non-zero $q$ the wavefunctions become up to signs those of the oscillator model based on the even-dimensional representations of $\mathfrak{u}(2)_\alpha$, see \cite{JSV2011}, for a specific parameter value in dimension $2j$. 
Recall that these correspond precisely to the even-dimensional representations of $\mathfrak{su}(2)_P$ obtained in Proposition~\ref{prop1}.

The described behaviour happens according to the following relations (for $q=1,2,\ldots,j$):
\[
\lim_{\gamma \to -1} \tilde R_{n}(\lambda(q);\gamma,0,j) = - \tilde R_{n-1}(\lambda(q-1);1,0,j-1)
\]
\[
\lim_{\delta \to -1} \tilde R_{n}(\lambda(q);0,\delta,j) = \tilde R_{n}(\lambda(q-1);0,1,j-1)
\]

We now look what happens to $\phi^{(c)}_n(q)$ for general $\tilde c$ when $j$ tends to infinity. 
This is done  by putting $q=j^{1/2} x$ to pass from a discrete position variable $q$ to a continuous position variable $x$ and taking the limit $j\rightarrow \infty$ of $j^{1/4} \phi^{(c)}_n(q)$. 
The actual computation is similar to the one performed in \cite{JSV2011,JSV2011b}, so we shall not give all details. 
The limit of the ${}_3F_2$ function in~\eqref{Phi-even} and \eqref{Phi-odd} is quite easy:
\begin{equation}
\lim_{j\rightarrow \infty} 
{\;}_3F_2\left(\myatop{-j^{1/2}x,j^{1/2}x,-n}{(1-\tilde c)/2,-j};1 \right) =
{\;}_1F_1\left(\myatop{-n}{(1-\tilde c)/2}; x^2 \right)= 
\frac{n!}{(a)_n} L_n^{(a-1)}(x^2),
\end{equation}
\begin{equation}
\lim_{j\rightarrow \infty}
{\;}_3F_2\left(\myatop{-j^{1/2}x+1,j^{1/2}x+1,-n}{(3-\tilde c)/2,-j+1};1 \right) =
{\;}_1F_1\left(\myatop{-n}{(3-\tilde c)/2}; x^2 \right) = 
\frac{n!}{(a+1)_n} L_n^{(a)}(x^2),
\end{equation}
where $a = (1-\tilde c)/2$ and $L_n^{(\alpha)}$ is a Laguerre polynomial~\cite{Koekoek,Temme}.

The final result is:
\begin{equation}
\lim_{j\rightarrow\infty} j^{1/4} \phi^{(c)}_{2n}(j^{1/2} x ) = 
(-1)^n \sqrt{\frac{n!}{\Gamma(a+n)}}\; |x|^{a-1/2} e^{-x^2/2} L_n^{(a-1)}(x^2),
\label{psi-even}
\end{equation}
\begin{equation}
\lim_{j\rightarrow\infty} j^{1/4} \phi^{(c)}_{2n+1}(j^{1/2} x ) = 
(-1)^n \sqrt{\frac{n!}{\Gamma(a+n+1)}}\; x |x|^{a-1/2} e^{-x^2/2} L_n^{(a)}(x^2).
\label{psi-odd}
\end{equation}
Note that for $\tilde c = 0$ or $a=1/2$, one indeed finds the canonical oscillator wavefunctions
\begin{equation}
\lim_{j\rightarrow \infty} j^{1/4} \phi^{(0)}_n(j^{1/2} x ) = \frac{1}{2^{n/2}\sqrt{n!}\pi^{1/4}} H_n(x) e^{-x^2/2},
\label{Hermite}
\end{equation}
where $H_n(x)$ are the common Hermite polynomials~\cite{Koekoek,Temme}.

The functions in~\eqref{psi-even} and \eqref{psi-odd} are familiar: they are in fact the wavefunctions $\Psi^{(a)}_n(x)$ of 
the parabose oscillator with parameter $a>0$ (see the appendix of \cite{JSV2011} for a summary). So we have: 
\begin{equation}
\lim_{j\rightarrow\infty} j^{1/4} \phi^{(c)}_{n}(j^{1/2} x ) = \Psi^{(a)}_n(x) \qquad(a=\frac{1-{\tilde c}}{2}).
\end{equation}
So the current model is an appealing model for a finite one-dimensional parabose oscillator with equidistant position spectrum. 
This also explains the shape of the discrete wavefunctions plotted in Figure~1.
For $-1<\tilde c<0$, the shape typically reproduces the continuous wavefunctions of the parabose 
oscillator with $\frac12<a<1$: see the plots for $\tilde c=-0.8$ and those for $a=0.9$ in Figure~2. 
For $0<\tilde c<1$, the shape of the wavefunctions is similar to those of the parabose oscillator with $0<a<\frac12$: compare the plots 
for $\tilde c=0.8$ with those for $a=0.1$ in Figure~2.

\section{Concluding remarks} 
\label{sec5}

Deformations or extensions of $\mathfrak{su}(2)$ or $\mathfrak{u}(2)$ as algebras underlying finite oscillator models have already been considered by one of us~\cite{JSV2011,JSV2011b}, so let us explain the difference with the algebra $\mathfrak{su}(2)_P$ appearing here.
For this, it is best to return to the classification of so-called dual Hahn doubles in~\cite{DoubleHahn}, where it is shown that three such doubles or pairs exist.
From~\cite[Propositions~1--3]{DoubleHahn} one can see that only the cases ``dual Hahn~I'' and ``dual Hahn~III'' can give rise to an equidistant position spectrum when used in a finite oscillator model.
The case ``dual Hahn~I'' involves the pair of polynomials $R_n(\lambda(x);\gamma,\delta,N)$ and $R_n(\lambda(x-1);\gamma+1,\delta+1,N-1)$, and the corresponding algebra constructed from the related tridiagonal matrices was determined in~\cite[eq.~(7.4)]{DoubleHahn}.
Comparing with $\mathfrak{su}(2)_P$, the relations \eqref{P}--\eqref{J0Jpm} remain the same, and~\eqref{JpJm} is of the form
\begin{equation}
[J_+,J_-]=2 J_0  + 2 (\gamma+\delta+1)J_0P - (2N+1)(\gamma-\delta)P + (\gamma-\delta)I. \label{7.4}
\end{equation}
Because of the appearance of $N$ and $N-1$ in the double, the matrices (and thus also the representations) exist in odd dimension $2N+1$ only;
furthermore the spectrum of the position operator consists of the values $0, \pm\sqrt{k(k+\gamma+\delta+1)}$ ($k=1,\ldots,N$).
For $\gamma=\delta\equiv\alpha$, \eqref{7.4} coincides with~\cite[eq.~(5)]{JSV2011b}, so this is what was called the $\mathfrak{su}(2)_\alpha$ extension in~\cite{JSV2011b}. The position spectrum is not equidistant. Moreover, due to the combination of terms in $J_0$ and $J_0P$ in the commutator of $J_+$ and $J_-$, this algebra cannot be rewritten as a special case of the Bannai-Ito algebra.

For $\delta=-\gamma-1$, \eqref{7.4} becomes 
\[
[J_+,J_-]=2 J_0+ (2\gamma+1)I   - (2N+1)(2\gamma+1)P 
\]
and after performing a shift for $J_0$, this relation is of the form~\eqref{JpJm}. So this is $\mathfrak{su}(2)_P$ (isomorphic to a special case of Bannai-Ito), the position spectrum is equidistant and this is the case (with odd dimensional representations) that was treated in the current paper.

The case ``dual Hahn~III'' involves the pair of polynomials $R_n(\lambda(x);\gamma,\delta,N)$ and $R_n(\lambda(x);\gamma+1,\delta-1,N)$, and the corresponding algebra constructed from the related tridiagonal matrices was determined in~\cite[eq.~(7.5)]{DoubleHahn}, with relation
\begin{equation}
[J_+,J_-]=2 J_0  + 2 (\gamma-\delta)J_0P - ((2N+2)(\gamma+\delta+1)+(2\gamma+1)(2\delta+1))P + (\gamma-\delta)I. \label{7.5}
\end{equation}
Because of the appearance of $N$ and $N$ in the polynomials of the double, the matrices (and representations) exist in even dimension $2N+2$ only; the spectrum of the position operator consists of the values $\pm\sqrt{(k+\gamma+1)(k+\delta+1)}$ ($k=0,\ldots,N$). So -- apart from a gap in the middle -- it is equidistant for $\gamma=\delta\equiv \alpha$, and then the above relation becomes
\[
[J_+,J_-]=2 J_0  - ((2N+2)(2\alpha+1)+(2\alpha+1)^2)P = 2 J_0  -(2\alpha+1)^2P - (2\alpha+1)CP,
\]
for some central element $C$. This was called the $\mathfrak{u}(2)_\alpha$ algebra in~\cite{JSV2011}. But since $C$ is a constant in a representation of the algebra, it can be considered as the $\mathfrak{su}(2)_P$ algebra with $-c=(2N+2)(2\alpha+1)+(2\alpha+1)^2$ in~\eqref{JpJm}. 
So $\mathfrak{su}(2)_P$ and $\mathfrak{u}(2)_\alpha$ are essentially the same, and the even dimensional representations of this algebra are the ones studied in~\cite{JSV2011}. 

\vskip 3mm
For the Lie algebra $\mathfrak{su}(2)$, there is of course the well known Schwinger boson realization.
In this realization, for a positive integer or half-integer $j$, the $2j+1$ basis vectors can be expressed as follows
\begin{equation}
|j,m\rangle   =\frac{ x^{j+m} y^{j-m} }{\sqrt{(j+m)!(j-m)!}},
\label{basisjm}
\end{equation}
and the $\mathfrak{su}(2)$ operators take the form
\begin{equation}
J_0=\frac12(x\partial_x - y \partial_y) ,\quad J_+ = x\partial_y,\quad J_- = y\partial_x.
\end{equation}
For the algebra $\mathfrak{su}(2)_P$, there exist similar reflection/differential operator realizations, one of which follows from the Bannai-Ito algebra realization and can be found in~\cite{BI}.
Since our basis elements~\eqref{P}--\eqref{JpJm} of $\mathfrak{su}(2)_P$ are closely related to the standard basis of $\mathfrak{su}(2)$, it is natural to expect other operator realizations.
These do indeed exist.

As a first possibility, consider the following operators, acting on functions $f(x,y)$ of two variables $x$ and $y$:
\begin{align}
T_x & = \partial_x + \frac{\mu}{x}(1-R_x) \nonumber \\
T_y & = \partial_y - \frac{\mu}{y}(1-R_x) .
\label{Ty1}
\end{align}
Herein, $R_x f(x,y) = f(-x,y)$. Note that $T_x$ is a Dunkl operator, but $T_y$ is not. 
Putting
\begin{equation}
J_0=\frac12(x\partial_x - y \partial_y+2\mu) ,\quad J_+ = xT_y,\quad J_- = yT_x, \quad P=R_x,
\label{su2Pxy1}
\end{equation}
it is easy to verify that the defining relations~\eqref{P} and~\eqref{J0Jpm} are satisfied.
For~\eqref{JpJm}, one finds:
\begin{equation}
[J_+,J_-] =  2J_0 - 2\mu P (1+x\partial_x + y \partial_y).
\end{equation}
So when acting on homogeneous polynomials in $x$ and $y$, like on the basis vectors~\eqref{basisjm}, the last relation coincides with~\eqref{JpJm} for $\mu=-\tilde c/2$.
For a proper action on homogeneous polynomials, one should take care of the factor $1/y$ in~\eqref{Ty1}: 
the action of $T_y$ on $x^{2j}$ should vanish. 
This is the case only for integer $j$-values, thanks to the factor $(1-R_x)$ in~\eqref{Ty1}.
Thus, the realization~\eqref{Ty1}--\eqref{su2Pxy1} is consistent with the basis realization~\eqref{basisjm} for $j$ integer only.
Note that on the space of homogeneous polynomials of degree $2j$, spanned by~\eqref{basisjm}, the action of $T_y$ does coincide with the action of a Dunkl operator $\partial_y - \frac{\mu}{y}(1-R_y)$, where $R_y f(x,y) = f(x,-y)$.

As a second possibility, let us take
\begin{align}
T_x & = \partial_x + \frac{\mu}{x}(1-R_x) \nonumber \\
T_y & = \partial_y + \frac{\mu}{y}(1+R_x) .
\label{Ty2}
\end{align}
and 
\begin{equation}
J_0=\frac12(x\partial_x - y \partial_y) ,\quad J_+ = xT_y,\quad J_- = yT_x, \quad P=-R_x.
\label{su2Pxy2}
\end{equation}
Once again, \eqref{P} and~\eqref{J0Jpm} are satisfied, and for~\eqref{JpJm} one finds:
\begin{equation}
[J_+,J_-] =  2J_0 + P \left((2\mu)^2+2\mu (1+x\partial_x + y \partial_y)\right).
\end{equation}
In this case, acting on homogeneous polynomials in $x$ and $y$ like on the basis vectors~\eqref{basisjm}, 
the last relation coincides with~\eqref{JpJm} for $c=(2\mu)^2+2\mu(2j+1)$ (in agreement with Remark~3).
Also here, one should take care of the factor $1/y$ in~\eqref{Ty2}, and the action of $T_y$ on $x^{2j}$ should vanish. 
This is now the case only for half-integer $j$-values, due to the factor $(1+R_x)$ in~\eqref{Ty1}.
The conclusion is similar: the realization~\eqref{Ty2}--\eqref{su2Pxy2} is consistent with the basis realization~\eqref{basisjm} for $j$ half-integer only.
For more fundamental examples in which such realizations with Dunkl operators play a role, see the Schwinger-Dunkl algebra $sd(2)$ in~\cite{GIVZ2013}.

\vskip 3mm
To summarize, in this paper we have developed a new and interesting model for a finite quantum oscillator.
This model preserves all the nice and essential properties of the original $\mathfrak{su}(2)$ model, in particular the equidistance of the position spectrum.
It has, however, an extra parameter $\tilde c$ that can be used to modify the shape of the discrete position (and momentum) wavefunctions.
The original interest in finite oscillator models comes mainly from optical image processing and signal analysis~\cite{Atak2005}.
In signal analysis on a finite number of discrete sensors or data points, one-dimensional finite oscillator models have been used in~\cite{Atak1994,Atak1997,Atak1999b}.
For such purposes, it is an advantage if the “sensor points” of the grid are uniformly distributed, according to the equidistant position spectrum of the model.
For our original Hahn oscillator in even dimensions~\cite{JSV2011} or in odd dimensions~\cite{JSV2011b}, this equidistance did not hold.
In the current model, based on a dual Hahn double, we do recover this important property of the spectrum (in odd dimensions).
We hope that the extra parameter $\tilde c$ opens the way to more sophisticated techniques in the analysis of signals.

The model presented here has the algebra $\mathfrak{su}(2)_P$ as underlying structure.
This algebra is an extension of $\mathfrak{su}(2)$ by $cP$, where $P$ which is not central but satisfies $P^2=1$ and either commutes or anticommutes with the standard basis elements of $\mathfrak{su}(2)$. 
We have shown that $\mathfrak{su}(2)_P$ is a special case of the general Bannai-Ito algebra.
For $\mathfrak{su}(2)_P$, we have classified all unitary finite-dimensional irreducible representations. These depend on the central element $c$.
Once the algebra and its representations have been analysed, the construction of the corresponding finite oscillator model is similar to that of~\cite{JSV2011}. 
The position wavefunctions are expressed in terms of dual Hahn polynomials (with different parameters for even and odd wavefunctions), and depend on the dimension of the representation ($2j+1$) and the parameter $\tilde c$ with $|\tilde c|<1$.
For $\tilde c=0$, the model and its wavefunctions coincide with the standard $\mathfrak{su}(2)$ finite oscillator model in terms of symmetric Krawtchouk polynomials~\cite{Atak2005}.
Symmetric Krawtchouk wavefunctions can interpreted as a finite-dimensional version of the canonical Hermite wavefunctions, to which they tend when the dimension paramater $j$ goes to infinity.
There is a similar interpretation here.
For $\tilde c\ne 0$, the wavefunctions can be seen as a finite-dimensional version (with equidistant spectrum) of the parabose wavefunctions.

\newpage

\newpage
\begin{figure}[htb]
\includegraphics{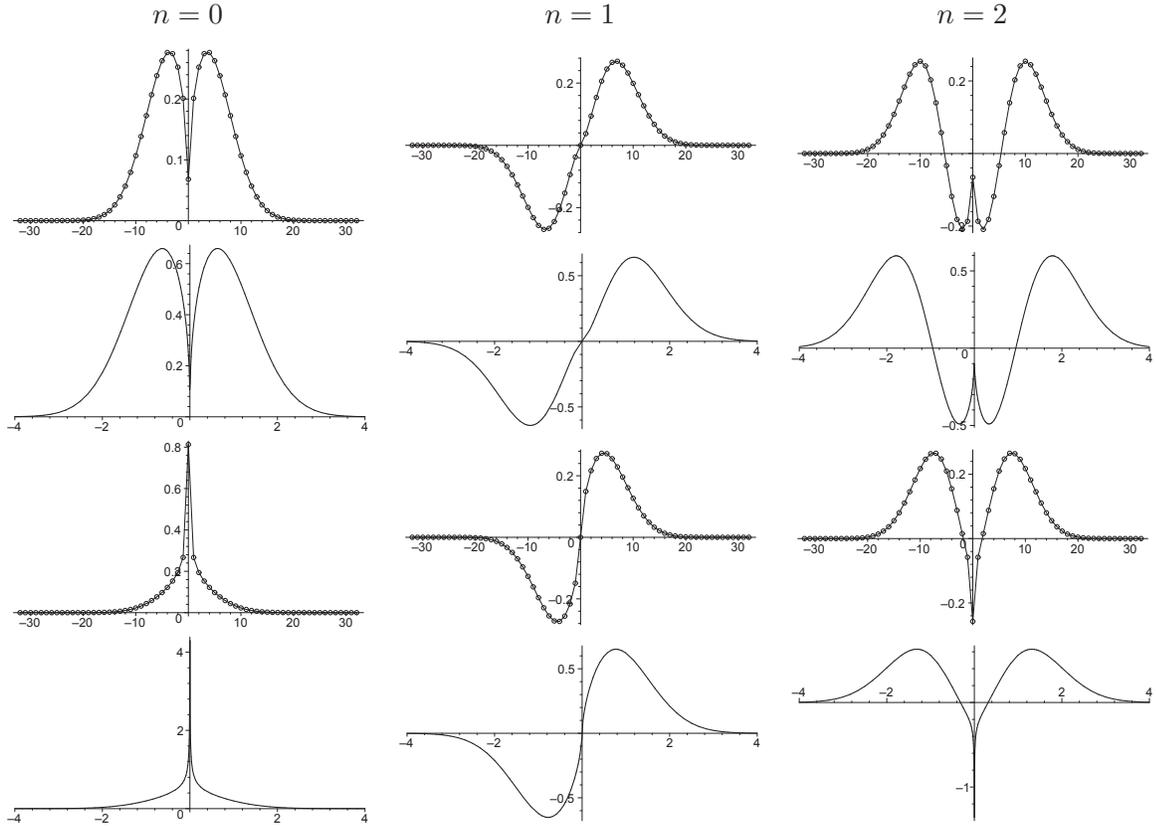}
\caption{
	Comparing the plots of the discrete wavefunctions $\phi^{(c)}_n(q)$ with the continuous wavefunctions $\Psi^{(a)}_n(x)$ of the parabose oscillator, for $n=0$ (left column), $n=1$ (middle column) and $n=2$ (right column).
	In the top row one finds $\phi^{(c)}_n(q)$ for $\tilde c=-0.8$, to be compared to the plots of $\Psi^{(a)}_n(x)$ in the second row for $a=0.9$.
	In the third row one finds $\phi^{(c)}_n(q)$ for $\tilde c=0.8$, to be compared to the plots of $\Psi^{(a)}_n(x)$ in the fourth row for $a=0.1$.
	}
	\label{fig2}
\end{figure}

\end{document}